\documentclass[pre,twocolumn,showpacs]{revtex4}
\usepackage{epsfig}

\newcommand{\g}[1]{{\bf {#1}}}

\begin{document}

\title{Local approximation for contour dynamics in effectively 
two-dimensional ideal electron-magnetohydrodynamic flows}
\author{V.~P. Ruban$^{1,2}$}
\email{ruban@itp.ac.ru}
\author{ S.~L. Senchenko$^{2,3}$}
\affiliation{$^{1}$L.D.Landau Institute for Theoretical Physics,
2 Kosygin Street, 119334 Moscow, Russia}
\affiliation{$^{2}$Optics and Fluid Dynamics Department, 
Ris\o ~National Laboratory, DK-4000 Roskilde, Denmark}
\affiliation{$^{3}$Danish Technical University, Department of Physics,
DK-2800 Lyngby, Denmark}

\date{\today}

\begin{abstract}
The evolution of piecewise constant distributions of a conserved
quantity related to the frozen-in canonical vorticity in effectively 
two-dimensional incompressible ideal EMHD flows is analytically investigated 
by the Hamiltonian method. The study includes the case of axisymmetric 
flows with zero azimuthal velocity component and also the case of
flows with the helical symmetry of vortex lines. 
For sufficiently large size of such a patch of the conserved quantity, 
a local approximation in the dynamics of the patch boundary is suggested, 
based on the possibility to represent the total energy as the sum of area 
and boundary terms. Only the boundary energy produces deformation of the shape 
with time. Stationary moving configurations are described.
\end{abstract}

\pacs{52.30.Cv}

\maketitle

\section{Introduction}

In this paper we consider a special class of vortical flows in plasma 
that correspond to the model of ideal electron magnetohydrodynamics (EMHD)
(see, e.g., \cite{KChYa1990, ABEKPSS1998, BSZCD1999, ACP2000} 
and references therein about EMHD and its applications).
In general, we deal with ideal EMHD-flows that are effectively 
two-dimensional (2D) and completely determined by a single function 
of two spatial coordinates and the time, namely with the usual planar flows, 
with the axisymmetric flows having the only azimuthal component of the 
magnetic field, and with the flows possessing the helical symmetry of 
frozen-in lines of the generalized vorticity. 
Our purpose here is to introduce a simplified analytical description 
adopted for the case when the function is piecewise constant 
(such distributions are usually called ``patches''), with a typical 
spatial scale between well separated the electron inertial length and 
the ion inertial length. In a narrow region near sharp boundary of a ``patch'' 
the density of the electric current is high, thus the boundary can be
considered as a dissipationless current sheet. From analytical viewpoint,
patches are the simplest examples of EMHD-flows with 
current sheets, in the sense that no other degrees of freedom are present 
in the system, except the shape of the patch.   
It is a well known fact that spontaneous formation of current sheets 
from initially smooth magnetic field configuration is a typical phenomenon 
in plasma dynamics, and the role of current sheets is very important. 
That is why it is interesting to investigate the problem of their motion, 
even in such highly idealized formulation as we do in this work.
As an explicit instance of fundamental physical phenomena tightly linked with 
our analytical study, we can indicate the fast magnetic field penetration 
into plasmas due to the Hall effect, since in some conditions the ideal EMHD 
is applicable to describe this phenomenon \cite{Fruchtman1991,SGFOO1996}.
More precisely, we focus on the effect of the electron inertia on the dynamics
of a narrow moving front between magnetized and unmagnetized plasma for those 
special highly symmetric configurations of the generalized vorticity field. 
It should be noted that on the mentioned scales the magnetic field is similar 
to the generalized vorticity field almost everywhere except the close vicinity 
of the sharp patch boundary, where the magnetic field is smoothed
on the electron inertial length.

As known, the EMHD model approximately describes the motion of the low-inertial 
electron component of plasma on sufficiently short scales 
(below the ion inertial length), while the much heavier ion component may be 
considered as motionless \cite{KChYa1990}.
Conditions for electrical quasi-neutrality of the plasma are assumed, which
imply that the macroscopic velocity $\g{v}(\g{r},t)$ of the electron flow 
should be sufficiently small and also all potential waves and oscillations 
of the plasma density should not be excited. Only vortical degrees of freedom
of the system are relevant in such circumstances. With these conditions, 
everywhere in the space the concentration $n(\g{r},t)$ of electrons is 
approximately equal to the prescribed concentration $N(\g{r})$ of ions, 
thus the density of flow of the electrons, 
$$
\g{j}(\g{r},t)\equiv n(\g{r},t)\g{v}(\g{r},t),
$$
is almost divergence free, $\mbox{div}\g{j}\approx 0$. The role 
of ions is thus reduced simply to compensation of the electric charge of 
electrons by providing the static neutralizing background. 
Such flows of the electron fluid create the divergence-free field 
of electric current density $-e\g{j}$, where $-e$ is the electron charge.
Hence, the quasi-stationary magnetic field 
\begin{equation}\label{magnetic_field}
\g{B}(\g{r},t)=-\frac{4\pi e}{c}\mbox{curl}^{-1}\g{j}(\g{r},t)
\end{equation}
should be taken into account when considering forces acting upon 
the electron fluid.

Important for our consideration is that in some cases dissipative effects
due to finite resistivity and/or viscosity 
may be neglected without dramatic loss of accuracy. 
So we study below the conservative EMHD model. 
A remarkable feature of the conservative hydrodynamic-type systems  
is that all of them possess an infinite number of integrals of motion related 
to a basic physical property of fluids, the relabeling symmetry 
(see, e.g., \cite{ZK97, R99, KR2000PRE, R2001PRE} and references therein
for a recent discussion). 
This symmetry manifests itself
as the freezing-in property of a divergence-free vector field
(the canonical vorticity field)
\begin{equation}\label{frozen-in_field}
\g{\Omega}(\g{r},t)=\mbox{curl}\left(\frac{\delta{\cal L}}{\delta\g{j}}\right),
\end{equation}
specified by the Lagrangian functional ${\cal L}\{n,\g{j}\}$ 
of a model and evolving with time accordingly to the equation
\begin{equation}\label{frozen-in_dynamics}
\g{\Omega}_t=\mbox{curl}[\g{v}\times\g{\Omega}].
\end{equation} 
For example, in the usual Eulerian hydrodynamics $\g{\Omega}(\g{r},t)$ is simply 
proportional to the velocity curl. As to the ideal EMHD, it is well known fact 
and we will see it below once more that frozen-in is the field
\begin{eqnarray}\label{EMHD_frozen-in_field}
\g{\Omega}(\g{r},t)&=&
\mbox{curl\,}m\g{v}(\g{r},t)-\frac{e}{c}\g{B}(\g{r},t) \nonumber\\
&=&\mbox{curl\,}m\g{v}+\frac{4\pi e^2}{c^2}\mbox{curl}^{-1}\g{j},
\end{eqnarray}
where $m$ is the electron mass. 

Below we deal with a spatially homogeneous ion background $N=\mbox{const}$,
thus the flows of the electron fluid are supposed to be incompressible,
$(\nabla\cdot\g{v})=0$.
There exist three classes (hereafter referred as $P,A,H$) 
of incompressible ideal EMHD flows having a special 
symmetry which allows one to describe the vorticity distribution in terms 
of a single scalar function $\omega$ depending on two spatial coordinates 
and the time. 
Besides the usual planar flows with 
\begin{equation}\label{typeP}
\g{\Omega}_P=\g{e}_z\omega(x,y,t),
\end{equation}
these effectively two-dimensional are the axisymmetric flows with zero 
azimuthal velocity component, 
\begin{equation}\label{typeA}
\g{\Omega}_{A}=[\g{e}_z\times\g{r}]\omega(z,q,t),
\end{equation}
where 
$$
q=(x^2+y^2)/2,
$$
and also the flows with the helical symmetry of the frozen-in vortex lines, 
\begin{equation}
\Omega^z_{H}=\omega(x\cos Kz+y\sin Kz,-x\sin Kz+y\cos Kz,t),
\end{equation}
\begin{equation}
\Omega^x_{H}=-Ky\Omega^z_{H},\qquad \Omega^y_{H}=Kx\Omega^z_{H},
\end{equation}
that are space-periodic along $z$-direction with the period $L^z=2\pi/K$.
What is important, in each of these three cases the evolution of 
the corresponding $\omega(u,v,t)$
is nothing else but only the transport of its level contours by 
2D incompressible flows. The general structure of the equations of motion 
for the function $\omega(u,v,t)$ is
\begin{equation}\label{transport_equation}
\omega_t+\Psi_v\omega_u -\Psi_u\omega_v =0,
\end{equation}  
with the stream-function $\Psi(u,v,t)$ being specified by the Hamiltonian 
functional ${\cal H}_\sigma\{\omega(u,v)\}$ of the model, depending also 
on the type of symmetry ($\sigma=P,A,H$),
\begin{equation}\label{Psi_definition}
\Psi=\frac{\delta{\cal H}_\sigma}{\delta\omega}.
\end{equation}
We will see later that the Hamiltonians ${\cal H}_\sigma\{\omega(u,v)\}$, 
corresponding to the ideal EMHD, take the quadratic form
\begin{equation}\label{Hamiltonian_general_form}
{\cal H}_\sigma\{\omega(u,v)\}=\frac{1}{2}\int 
\omega(u,v)\hat G_\sigma\omega(u,v)du\,dv,
\end{equation}
where the nonlocal linear operators $\hat G_\sigma$ possess smoothing 
properties like the usual two-dimensional $\Delta^{-1}$-operator. 
Therefore the flows with discontinuous piecewise constant distributions 
of the function $\omega(u,v)$ are possible. From 
Eqs.(\ref{transport_equation}-\ref{Psi_definition}) 
it follows that the shape 
$\{u(\xi,t),v(\xi,t)\}$ of boundary of such a patch with 
$\omega=\omega_0=\mbox{const}$ (here $\xi$ is an arbitrary 
longitudinal parameter),  evolves in accordance with the equation
\begin{equation}\label{patch_dynamics_general_form}
u_t v_\xi -u_\xi v_t=\frac{1}{\omega_0}\frac{\partial}{\partial \xi}\left[
\frac{v_\xi(\delta{\cal H}^*/\delta u)-
u_\xi(\delta{\cal H}^*/\delta v)
}{u_\xi^2+v_\xi^2}\right],
\end{equation}
where ${\cal H}^*\{u(\xi),v(\xi)\}$ is the energy of the patch 
expressed through its shape.
The purpose of this work is to study the motion of the patches 
in $(u,v)$-plane for all three kinds of the geometric
symmetry. Wee will extensively use the fact that in EMHD the kernels 
$G_\sigma(u_1,v_1; u_2,v_2)$ of the operators are exponentially small if 
the distances between $(u_1,v_1)$ and $(u_2,v_2)$ are much longer 
than an internal length parameter of the problem, the inertial electron skin 
depth $d=(mc^2/4\pi e^2n)^{1/2}$. For comparatively large patches, 
this property makes possible to single out the local boundary term in 
the Hamiltonian, which is responsible for the evolution of the patch shape, 
while the main area term in the approximate local Hamiltonian only results in 
a uniform motion of the patch without changing the shape.

\section{Canonical formalism for EMHD}

For convenience and self-consistency, now we reproduce briefly derivation of 
the EMHD equations, following the canonical formalism adopted for fluids as 
described in Refs. \onlinecite{R99,R2001PRE}.
As the start point, let us consider the microscopic Lagrangian of a 
system of electrically charged point particles that can be written up to 
the second order on $v/c$, as it is given in the famous book by Landau 
and Lifshitz \cite{LL2},
\begin{eqnarray}
&&{\cal L}_{\mbox{\small micro}}=\sum_a\frac{m_a\g{v}_a^2}{2}
-\frac{1}{2}\sum_{a\not =b}\frac{e_a e_b}{|\g{r}_a-\g{r}_b|}
+\sum_a\frac{m_a\g{v}_a^4}{8c^2}\nonumber\\
&&+\frac{1}{4c^2}\sum_{a\not =b}\frac{e_a e_b}{|\g{r}_a-\g{r}_b|}
\{\g{v}_a\cdot\g{v}_b+(\g{v}_a\cdot\g{n}_{ab})(\g{v}_b\cdot\g{n}_{ab})\},
\label{Lmicro}
\end{eqnarray}
where $\g{r}_a(t)$ are the positions of the point charges $e_a$, 
$\g{v}_a(t)\equiv\dot\g{r}_a(t)$ are their velocities, 
$\g{n}_{ab}(t)$ are the unit vectors in the direction between $e_a$ and $e_b$.
The first double sum in Eq.(\ref{Lmicro}) corresponds to the electrostatic 
interaction, while the second double sum 
describes the magnetic interaction via quasi-stationary magnetic field. 
It is very important that for a system with macroscopically large number
of particles the magnetic energy can be of the same order (or even larger) as
the kinetic energy given by the first ordinary sum in Eq.(\ref{Lmicro}), 
while the terms of the fourth order on the velocities are often negligible. 
Generally speaking, a large part of plasma physics is governed by this 
Lagrangian, at least in the cases when the velocities of particles are 
non-relativistic
and the free electro-magnetic field is not excited significantly. Obviously,
different physical problems need different procedures of macroscopic
consideration of this system. The most accurate (and the most complicated) 
would be a kinetic description. However, for our purposes it is sufficient
to apply more simple and naive procedure of the hydrodynamical averaging,
that gives less accurate description of the system in terms of the 
concentration $n({\bf r},t)$ of electrons  and the density $\g{j}(\g{r},t)$
of their flow, that satisfy the continuity equation
\begin{equation}\label{contin}
\frac{\partial n}{\partial t}+\mbox{div}\g{j}=0. 
\end{equation} 
Neglecting all dissipative processes that take place due to collisions
of the particles (though on this step we strongly reduce applicability of the 
following conservative EMHD model), and considering the ions as 
macroscopically motionless, we derive from Eq.(\ref{Lmicro})
the following Lagrangian functional of the electron fluid on the 
given static ion background with the macroscopic concentration $N(\g{r})$:
\begin{eqnarray} \label{L_e}
&&{\cal L}_{e}\{n,\g{j}\}=\int\!d\g{r}\Big[m\frac{\g{j}^2}{2n}
-\varepsilon(n)\Big]\nonumber\\
&&-\frac{e^2}{2}\!\int\!\!\int \!\frac{d\g{r}_1d\g{r}_2}{|\g{r}_1-\g{r}_2|}
\{n(\g{r}_1)-N(\g{r}_1)\}\{(n(\g{r}_2)-N(\g{r}_2)\}\nonumber\\
&&+\frac{e^2}{4c^2}\!\int\!\!\int \!\frac{d\g{r}_1 d\g{r}_2}
{|\g{r}_1-\g{r}_2|}\Big[\g{j}(\g{r}_1)\cdot\g{j}(\g{r}_2)\nonumber\\
&&\qquad\qquad
+\frac{(\g{j}(\g{r}_1)\cdot \{\g{r}_1-\g{r}_2\})
(\g{j}(\g{r}_2)\cdot \{\g{r}_1-\g{r}_2\})}{|\g{r}_1-\g{r}_2|^2}\Big]
,
\end{eqnarray}
where the internal energy $\varepsilon(n)$ 
of the electron fluid takes into account energy of
the thermal disordered motion of electrons and also the 
microscopic-scale-concentrated part of the electro-magnetic energy. 
It should be kept in mind that $\varepsilon(n)$ depends also on the specific 
entropy but we suppose the flows isentropic. 

Since we are interested below in relatively 
slow vortical motion of the electron fluid, when all possible potential waves 
and oscillations are not excited significantly, so the quasi-neutrality 
condition $n\approx N$ is well satisfied, it is possible to neglect the 
second line in the expression (\ref{L_e}), as well as variation of 
$\varepsilon(n)$. Thus, for the EMHD, which describes this 
special dynamical limit of the system (\ref{L_e}), we have the Lagrangian
\begin{equation}\label{L_EMHD}
{\cal L}_n\{\g{j}\}=\int\left[m\frac{\g{j}^2}{2n}+
\frac{4\pi e^2}{c^2}\frac{(\mbox{curl}^{-1}\g{j})^2}{2}\right]d\g{r},
\end{equation}
where 
the concentration $n(\g{r})$ is a prescribed function, 
so the density $\g{j}$ of flow is divergence-free,
$\mbox{div}\g{j}=0. $

In general, the equation of motion for the divergence-free flux field $\g{j}$ 
has the form (compare with \cite{R2001PRE})
\begin{equation}\label{generalizedEuler}
\frac{\partial}{\partial t}\mbox{curl}\left(\frac{\delta 
{\cal L}\{\g{j}\}}{\delta\g{j}}\right)
=\mbox{curl}\left[\frac{\g{j}}{n}\times
\mbox{curl}\left(\frac{\delta {\cal L}\{\g{j}\}}{\delta\g{j}}
\right)\right].
\end{equation}
In the particular case for EMHD this yields
\begin{eqnarray} 
&&\frac{\partial}{\partial t}
\left(\mbox{curl\,}\frac{m\g{j}}{n} +\frac{4\pi e^2}{c^2}
\mbox{curl}^{-1}\g{j}\right)
\nonumber\\
&&\qquad =\mbox{curl}\left[\frac{\g{j}}{n}\times
\left(\mbox{curl\,}\frac{m\g{j}}{n} +\frac{4\pi e^2}{c^2}
\mbox{curl}^{-1}\g{j}\right)\right].\label{EMHDequation}
\end{eqnarray}
One can consider the canonical vorticity field 
\begin{equation}\label{Omega_definition}
\g{\Omega}\equiv\mbox{curl}\left(\frac{\delta {\cal L}_n}{\delta\g{j}}\right)
\end{equation}
and the Hamiltonian functional
\begin{equation}\label{H_definition}
{\cal H}\{\g{\Omega}\}\equiv\left\{\int 
\left(\frac{\delta {\cal L}_n}{\delta\g{j}}\cdot\g{j} \right)d\g{r}-{\cal L}_n
\right\}\Big|_{\g{j}=\g{j}\{\g{\Omega},n\}}.
\end{equation}
Using the relation 
\begin{equation}\label{j_Omega}
\g{j}=\mbox{curl}\left(\frac{\delta {\cal H}}{\delta\g{\Omega}}\right),
\end{equation}
it is possible to rewrite the equation of motion (\ref{generalizedEuler}) 
in the form
\begin{equation}\label{frozen-in_Hamiltonian_dynamics}
\g{\Omega}_t=\mbox{curl}\left[\mbox{curl}\left(\frac{\delta {\cal H}}
{\delta\g{\Omega}}\right)\times\frac{\g{\Omega}}{n}\right],
\end{equation}
that emphasizes the freezing-in property of the canonical vorticity.

Hereafter we consider the case $n=\mbox{const}$ and use dimensionless 
variables, with all length scales normalized to the ion inertial length. 
The expression for the EMHD Hamiltonian can be written then in the 
following form
\begin{equation}\label{EMHD_3D_Hamiltonian}
{\cal H}\{\g{\Omega}\}=\frac{1}{2}\int\!\int G(|\g{r}_1-\g{r}_2|)
(\g{\Omega}(\g{r}_1)\cdot\g{\Omega}(\g{r}_2) )
d\g{r}_1d\g{r}_2,
\end{equation}
with the Green's function
\begin{equation}\label{EMHD_Green_function}
G(r)=\frac{\exp(-r/\lambda)}{4\pi\lambda^2 r},
\end{equation}
where $\lambda=\sqrt{m/M}$ is a small parameter, the ratio of the electron
inertial length to the ion inertial length. It should be noted that the 
EMHD model in this simplest form, without taking into account the ion motion, 
is applicable only on scales below the ion inertial length. A more accurate 
model differs by another Green's function in the double integral 
(\ref{EMHD_3D_Hamiltonian}): instead of the expression 
(\ref{EMHD_Green_function}) with effectively finite radius of interaction,
one has to use the modified Green's function, 
with infinite radius of interaction
(see, e.g., \cite{physics/0110023} for more detail)
\begin{equation}\label{modified_EMHD_Green_function}
\tilde G(r)=\frac{1}{4\pi}\left(
\frac{\exp(-r/\lambda)}{\lambda^2 r}+\frac{1}{r}\right).
\end{equation}
However, in this work we deal with the function (\ref{EMHD_Green_function}),
since our goal is to construct a local approximate model. Thus, a typical
size $L$ of vortex structures must be in the limits 
$\lambda\ll L \ll 1$, in order the contribution to the Hamiltonian
from the second term in Eq.(\ref{modified_EMHD_Green_function}) 
to be small in comparison with the contribution from the first one.

\section{Long-scale local approximation in contour dynamics}

\subsection{Axisymmetric flows}
\begin{figure}
\begin{center}
  \epsfig{file=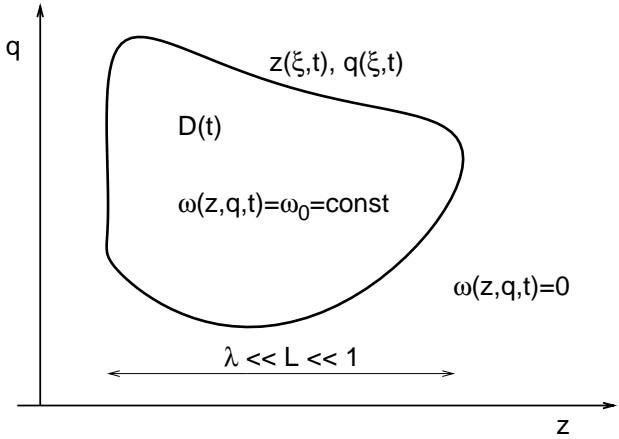,width=85mm}
\end{center}
\caption{\small Sketch of generalized vorticity patch in axisymmetric flows.} 
\label{patch}
\end{figure}
Let us consider effectively 2D flows, for instance, 
the axisymmetric flows (\ref{typeA}). The dynamics of the function 
$\omega(z,q,t)$ is determined by Eqs.(\ref{transport_equation}--
\ref{Psi_definition}), where ${\cal H}_{A}
\{\omega\}\!=\!(1/2\pi){\cal H}\{\omega[\g{e}_z\!\times\!\g{r}]\}$. 
For long-scale flows, the Green's function (\ref{EMHD_Green_function})
in the double integral
(\ref{EMHD_3D_Hamiltonian}) is almost the same as the $\delta$-function. 
Therefore, large vortex patches like that shown in the Fig.\ref{patch}, 
with constant $\omega$ and 
sharp boundaries, can be approximately 
described by a local Hamiltonian broken onto two parts --- the bulk energy 
$$
{\cal H}_{A}^D=\frac{1}{2}\int_D \g{\Omega}^2 rdr\,dz
=\omega_0^2\int_D q\,dq\,dz ,
$$
and a "surface" energy originated by the effect of non-locality near the 
boundary.
In leading order the boundary term can be calculated as if we have locally
a 1D configuration with the jump $\Omega=\omega_0 \sqrt{2q}$ in the vorticity 
field. In such a 1D case, the additional energy  per unit area of the boundary 
is simply 
$$
-\frac{\Omega^2}{2\cdot 2}\int_0^\infty e^{-\zeta/\lambda}d\zeta
=-\frac{\Omega^2\lambda}{4},
$$
that gives the surface energy of the patch in the axisymmetric flows,
\begin{eqnarray}
{\cal H}_{A}^{\partial D}=-\frac{\lambda\omega_0^2}{4}\oint_{\partial D} 
r^2\cdot r\sqrt{(dz)^2+(dr)^2}
\nonumber\\
= -\frac{\lambda\omega_0^2}{2}\oint_{\partial D} q\sqrt{2q (dz)^2+(dq)^2}.
\end{eqnarray}

Let us for simplicity take $\omega_0=1$. Using the explicit expression 
for the Hamiltonian functional 
${\cal H}_A^{*}\{z(\xi),q(\xi)\}={\cal H}_{A}^D+{\cal H}_{A}^{\partial D}$,
\begin{equation}\label{local_H_A}
{\cal H}_A^{*}\{z(\xi),q(\xi)\}=\oint  q z q_\xi d\xi \, 
-\frac{\lambda}{2}\oint  q\sqrt{2qz_\xi^2+q_\xi^2}d\xi,
\end{equation}
we obtain from Eq.(\ref{patch_dynamics_general_form}) the following
equation of motion, which does not depend on choice of the
longitudinal parameter $\xi$:
\begin{equation}\label{local_eq_motion}
z_t q_\xi-z_\xi q_t=q_\xi+\lambda\frac{\partial}{\partial\xi}
\left[\frac{1}{q_\xi}\frac{\partial}{\partial\xi}
\left(\frac{q^2z_\xi}{\sqrt{2qz_\xi^2+q_\xi^2}}\right)
\right].
\end{equation}
This local nonlinear equation is one of the main results of present work.
It approximates the nonlocal contour dynamics in axisymmetric EMHD-flows if 
a typical longitudinal scale $L$ (in $(r,z)$-plane) 
of the contour satisfies the condition
$\lambda\ll L \ll 1$.

Now we study stationary moving (along $z$-axis) configurations.
Let the shape of a patch $\omega_0=1$ in axisymmetric flow be given (locally)
by the function $z(q,t)$ (this implies the fixed choice of the longitudinal
parameter in Eq.(\ref{local_eq_motion}),  $q=\xi$). 
Then the corresponding equation of motion is 
\begin{equation}\label{z(q)_motion}
z_t(q)=1+\lambda\frac{\partial^2}{\partial q^2}
\left(\frac{q^2z_q}{\sqrt{1+2qz_q^2}}\right).
\end{equation}
Let us consider stationary moving solutions, $z_t=1+2\lambda C$, where
$C=\mbox{const}$. The shape of the boundary in such case is determined 
by the ordinary differential equation
\begin{equation}\label{z(q)_stationary}
\frac{\partial^2}{\partial q^2}\left(\frac{q^2z_q}{\sqrt{1+2qz_q^2}}\right)=2C,
\end{equation}
that can be easily integrated:
\begin{equation}\label{z(q)_stat_solution_ABC}
z=\int^q dq\frac{(Cq^2+\tilde a q+\tilde b)}
{\sqrt{q^4-2q(Cq^2+\tilde a q+\tilde b)^2}},
\end{equation}
with some constants $\tilde a, \tilde b, C$. 
To simplify further analysis it is convenient to 
make the rescaling
$$
q \mapsto \alpha q, \ \ \ z \mapsto  \sqrt{\alpha}z,
 \ \ \ \alpha=\frac{1}{2C^2}.
$$
In these new coordinates the stationary solutions
depend only on two parameters,
\begin{equation}\label{z(q)_stat_solution}
z=\frac{1}{\sqrt{2}}\int^{q} dq \frac{q^2+aq+b}{\sqrt{q^4-q(q^2+aq+b)^2}}
\label{two_param}.
\end{equation} 
However, in general case the constants $a$ and $b$ may not take arbitrary 
values, since the expression under the square root in the denominator must 
be positive in some range of (positive) $q$. 
Another restriction is that self-intersections of the curves are forbidden.
\begin{figure}
\begin{center}
  \epsfig{file=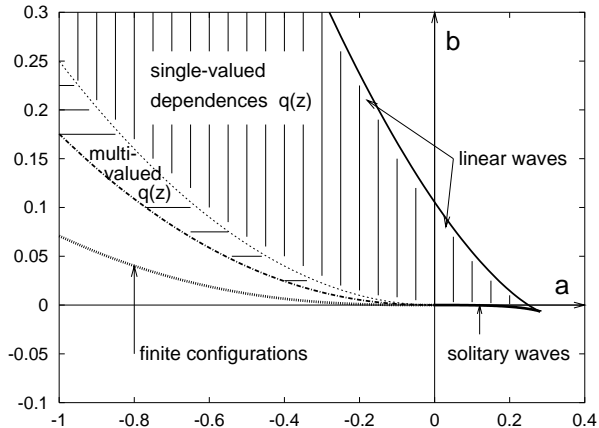,width=85mm}
\end{center}
\caption{\small The parametric plane for axisymmetric stationary solutions.} 
\label{par_plane}
\end{figure}
The parametric $(a,b)$-plane is shown in Fig.\ref{par_plane}. 
First, there exist $z$-periodic configurations that  occupy
2D region in the parametric plane and correspond to 
axisymmetric current channels with a sharp crimped surface. 
On different pieces of the boundary of that region we have different types
of stationary solutions. Partially the boundary is determined by the 
parametrically given curve (the thick solid line in Fig.\ref{par_plane})
\begin{equation}\label{parametric_curve}
a(\tau)=-2\tau(\tau-3/4), \qquad b(\tau)=\tau^3(\tau-1/2). 
\end{equation}
If the parameter $\tau$ is within the limits $3/8<\tau<+\infty$, then
the solutions near the curve look like  
linear sinusoidal waves with vanishing amplitude.
If $ 0<\tau<3/8$, then $z$-period of the current channels tends to infinity,
and we have highly nonlinear solitary waves. 
For the third piece of the boundary, where self-touch of the current channels 
with multi-valued dependences $q(z)$ takes place, 
we do not have a simple analytical expression, however it can be found
numerically. 
Second, besides the infinitely long current channels, 
there exist finite configurations (cross-sections of magnetic rings).
For finite configurations the curves in $(z,q)$-plane must be closed, 
thus in this case the integral (\ref{z(q)_stat_solution}) between the two 
corresponding zeros of the denominator must be equal to zero, that gives a
relation between $a$ and $b$ (the numerically calculated separate curve in 
Fig.\ref{par_plane}). Though the piece of the parametric curve 
(\ref{parametric_curve}) with 
$-\infty<\tau<0$  corresponds to nothing in the parametric plane, 
but the line of finite configurations approaches it from below, as $a\to 0$. 

\begin{figure}
\begin{center}
  \epsfig{file=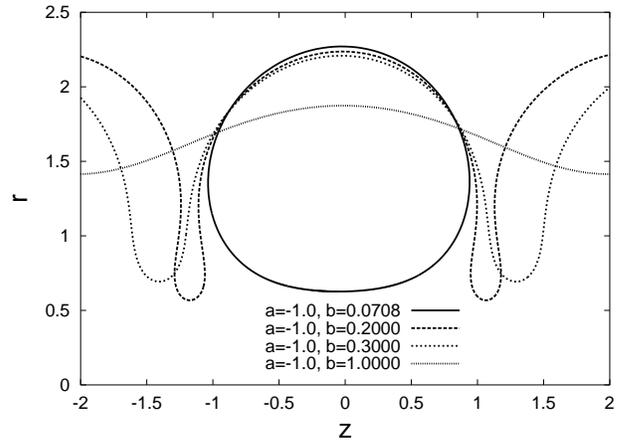,width=85mm}
\end{center}
\caption{\small 
Examples of axisymmetric solutions: crimped current channels 
and a magnetic ring.} 
\label{examples}
\end{figure}

\begin{figure}
\begin{center}
  \epsfig{file=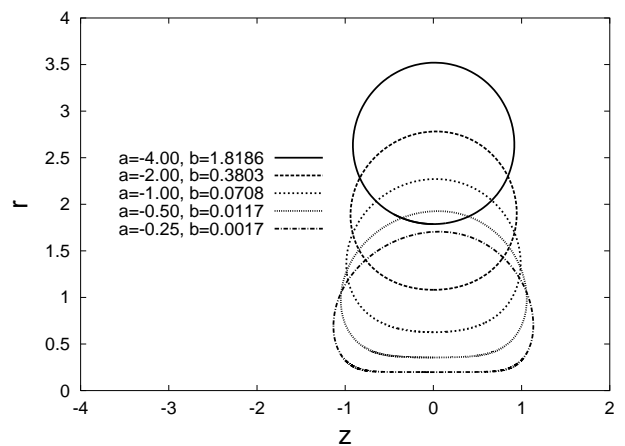,width=85mm}
\end{center}
\caption{\small Cross-sections of magnetic rings.} 
\label{ovals}
\end{figure}

Examples of appropriate solutions
are presented in Fig.\ref{examples} and Fig.\ref{ovals}. 
Strictly speaking, only the magnetic rings can satisfy the condition 
$ L\ll 1$. For the infinitely long current channels, applicability 
of the local model is not so well justified as for the rings. 

In the limit, when $a\to 0$, 
the ``upper" part of the contour shape is the parabola 
when being viewed in $(q,z)$-plane 
(the half-circle in $(r,z)$-plane),
$$
(z^2+r^2)=2.
$$
For the rings, the tendency towards this shape is clearly 
seen in the Fig.\ref{ovals}, as $|a|$ decreases.

\subsection{Helical flows}

Analogously, a patch $\omega=1$ in the helical flows can be investigated. 
For this case the bulk energy is
\begin{eqnarray}
{\cal H}_{H}^D&&=\frac{1}{2}\int_D [1+K^2(u^2+v^2)]du\, dv\nonumber\\
&&=\frac{1}{2}\oint_{\partial D}udv+\frac{K^2}{2}\oint_{\partial D}
(uv^2+u^3/3)dv,\label{H_bulk_Hamiltonian}
\end{eqnarray}
while the boundary energy can be expressed as follows:
\begin{equation}\label{III_boundary_Hamiltonian}
{\cal H}_{H}^{\partial D}=
-\frac{\lambda}{4}\oint_{\partial D}[1+K^2\g{w}^2]
\sqrt{(d\g{w})^2+K^2(\g{w}\cdot d\g{w})^2}
\end{equation}
Here the square root is proportional to the area element of the helical tube 
surface. 

Note that the planar flows (\ref{typeP}) are included into this consideration, 
since they correspond to the case $K=0$. The boundary term in that case is 
simply proportional to the length of the contour that results in the integrable 
dynamics \cite{GP1991,GP1992} --- the evolution of the boundary 
curvature $\kappa(l,t)$, where $l$ is the arc-length along the contour, 
is determined by the modified KdV equation,
\begin{equation}
\kappa_t=\frac{\lambda}{4}
\left(\kappa_{lll}+\frac{3}{2}\kappa^2\kappa_l\right).
\end{equation}

\begin{figure}
\begin{center}
  \epsfig{file=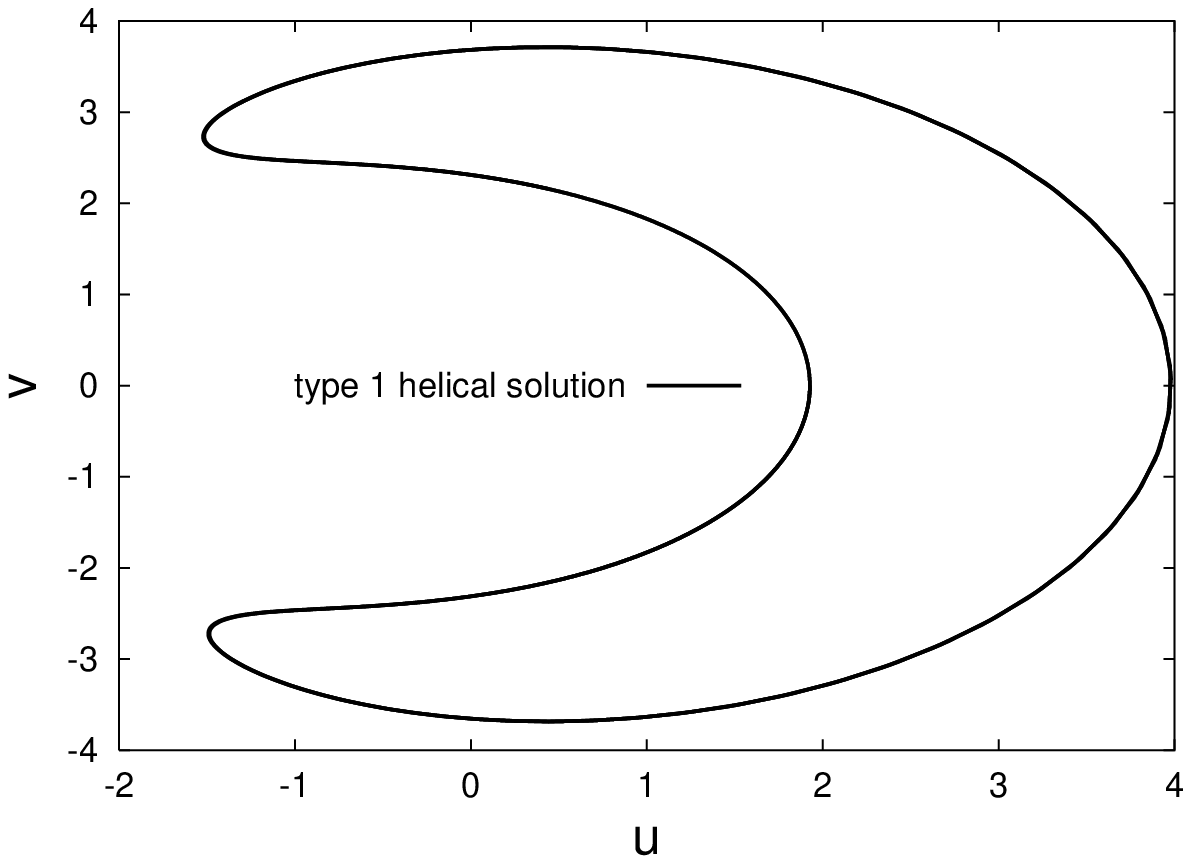,width=85mm}
  \epsfig{file=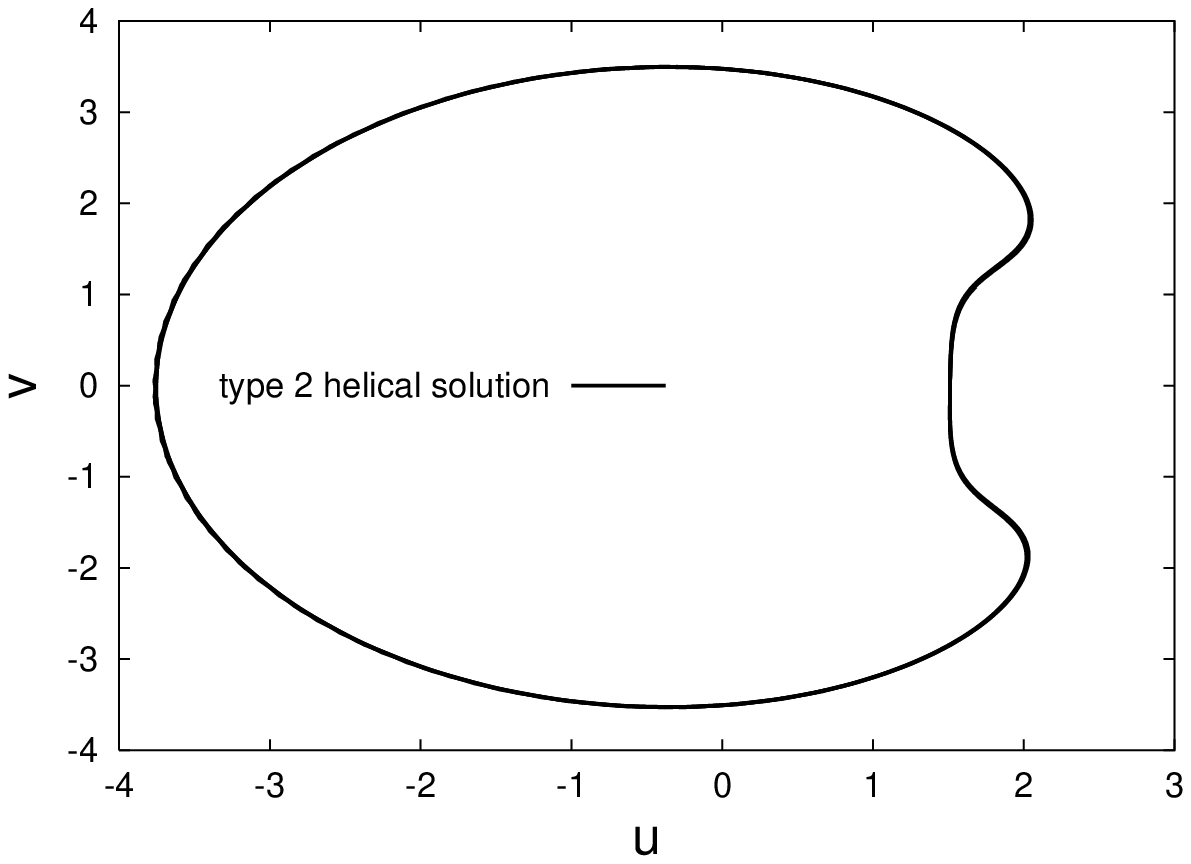,width=85mm}
\end{center}
\caption{\small 
Examples of stationary rotating patches in helical flows.} 
\label{helical}
\end{figure}

In the case $K\not =0$, 
the term ${\cal H}_{H}^D$ just produces the uniform rotation of the contour 
with the angular velocity $\dot\varphi_0=-K^2$ around the origin. 
The shape evolution generated by ${\cal H}_{H}^{\partial D}$ 
is more convenient for analytical study in the polar coordinates
(note the relation $\partial(u,v)/\partial(q,\varphi)=1$)
\begin{equation}
u=\sqrt{2q}\cos\varphi, \qquad v=\sqrt{2q}\sin\varphi,
\end{equation}
since the expression for ${\cal H}_{H}^{\partial D}\{\varphi(q)\}$ 
does not contain $\varphi(q)$ itself, but only $\partial_q\varphi(q)$:
\begin{equation}
{\cal H}_{H}^{\partial D}=
-\frac{\lambda}{4}\int dq[1+2K^2 q]
\sqrt{1/2q+2 q\varphi_q^2+K^2}
\end{equation}
The equation of motion is
\begin{eqnarray}
\varphi_t&&=-\frac{\partial}{\partial q}
\left(\frac{\delta ({\cal H}_{H}^D+{\cal H}_{H}^{\partial D})}
{\delta \varphi}\right)\nonumber\\
&&=-K^2-\frac{\lambda}{4}\frac{\partial^2}{\partial q^2}
\left(\frac{2q\varphi_q[1+2K^2 q]}
{\sqrt{1/2q+2 q\varphi_q^2+K^2}}
\right).\label{helical_eq_motion}
\end{eqnarray}
For stationary rotation, when $\varphi_t=-K^2-C\lambda$,we have
\begin{equation}
\varphi(q)=\int^q\frac{dq}{q}\frac{(Cq^2+Bq+A)\sqrt{1+2K^2q}}
{\sqrt{2q(1+2K^2q)^2-4(Cq^2+Bq+A)^2}}
\end{equation}
Here also a right choice of the constants $A,B,C$ must satisfy the condition 
for the curve to be closed and do not have self-intersections  
(see Fig.\ref{helical}).

\section{Discussion}

Thus, a local approximation for a class of effectively 2D flows 
in the ideal EMHD, based on the Hamiltonian formulation of the problem,
has been suggested for the nonlinear contour dynamics. Stationary moving
configurations with current sheets have been analytically (in integral form) 
found within this approximation. They describe propagation of magnetic rings, 
traveling waves on a crimped surface of current channels, including solitary
waves, and rotating helical magnetic structures. Of course, our results
are not able to explain all the important phenomena in EMHD.
A payment for the relative 
simplicity of our local models has been the general impossibility to consider 
in such a manner short-scale ($\sim\lambda$) contour perturbations,  
though their dynamics may have influence on the problem of 
stability of the stationary moving configurations. Also, the question about
evolution of 3D perturbations of the magnetic rings, current channels, 
and/or helical structures cannot be answered in such very simplified 
description of the flows by moving curves in a plane. 
But we believe our approach useful since it develops an insight in 
qualitative theoretical understanding of such complicated
nonlocal theory as the EMHD. For instance, we do not see another simple 
and compact way how to derive equations of motion for current sheets, 
analogous to Eqs.(\ref{local_eq_motion}) and (\ref{helical_eq_motion}).

\begin{acknowledgments}
These investigations were supported by the  Danish Graduate School 
in Nonlinear Science and by the INTAS (grant No. 00-00292).
The work of V. R. was supported also by RFBR (grant No. 00-01-00929),
by the Russian State Program of Support of the Leading Scientific Schools 
(grant No. 00-15-96007), and by the Science Support
Foundation, Russia.
\end{acknowledgments}

\end{document}